\documentclass[prb,amsmath,amssymb,twocolumn,longbibliography]{revtex4-2}
\usepackage{graphicx}
\usepackage{dcolumn}
\usepackage{bm}
\usepackage{placeins}
\usepackage{rotating}
\usepackage{multirow}
\usepackage[dvipsnames,usenames]{xcolor}
\usepackage[english]{babel}
\usepackage{amssymb}
\usepackage{amsmath}
\usepackage[utf8]{inputenc}
\usepackage[normalem]{ulem}
\usepackage{float}
\usepackage{siunitx}

\newcommand{\br}{\mathbf{r}}
\newcommand{\be}{\begin{equation}}
\newcommand{\ee}{\end{equation}}
\newcommand{\bea}{\begin{eqnarray}}
\newcommand{\eea}{\end{eqnarray}}
\newcommand{\nn} {\nonumber}

\newcommand{\Tr}{ {\rm Tr} \, }

\def\a{\alpha}
\def\b{\beta}

\def\G{\Gamma}
\def\d{\delta}
\def\D{\Delta}

\def\ve{\varepsilon}

\def\s{\sigma}
\def\S{\Sigma}

\def\vf{\varphi}

\def\w{\omega}

\def\bra{\langle}
\def\ket{\rangle}

\def\xc{{\rm xc}}

\def\Tr{{\rm Tr}\,}

\def\br{\mbox{\boldmath $r$}}

\graphicspath{{Figures/}}

\usepackage{soul}
\begin{document}

\title{Self-consistent random phase approximation and optimized \\hybrid functionals for solids}
\author{Thomas Pitts} 
\affiliation{Sorbonne Universit\'e, MNHN, UMR CNRS 7590, IMPMC, 4 place Jussieu, 75005 Paris, France}
\author{Damian Contant} 
\affiliation{Sorbonne Universit\'e, MNHN, UMR CNRS 7590, IMPMC, 4 place Jussieu, 75005 Paris, France}
\author{Maria Hellgren}
\affiliation{Sorbonne Universit\'e, MNHN, UMR CNRS 7590, IMPMC, 4 place Jussieu, 75005 Paris, France}
\date{\today}
\begin{abstract}
The random phase approximation (RPA) and the $GW$ approximation share the same total energy functional 
but RPA is defined on a restricted domain of Green's functions determined by a local Kohn-Sham (KS) potential. 
In this work, we perform self-consistent RPA calculations by optimizing the local KS potential through the optimized effective potential equation. We study a number of solids (C, Si, BN, LiF, MgO, TiO$_2$), and find in all cases a lowering of the total energy with respect to non-self-consistent RPA. We then propose a variational approach to optimize parameter-dependent hybrid functionals based on the minimization of the RPA total energy with respect to the fraction of exact exchange used to generate the input KS orbitals. We show that this scheme leads to hybrid functionals with a KS band structure in close agreement with RPA, and with lattice constants of similar accuracy as within RPA. Finally, we evaluate $G_0W_0$ gaps using RPA and hybrid KS potentials as starting points. Special attention is given to TiO$_2$, which exhibits a strong starting-point dependence.
\end{abstract}
\maketitle
\section{Introduction}
The random phase approximation (RPA) within Kohn-Sham (KS) density functional theory (DFT) has become an important approach in chemistry and materials science, owing to its ability to accurately handle complex systems with mixed van der Waals/covalent bonds \cite{Kresse2008,Bredow2008,Galli2009,Scheffler2009,Kresse2009,Kresse2011,Thygesen2011,SchefflerKim2012,SchefflerRen2012,Furche2012,Kresse2017,Tew2019,CazorlaGould2019,Ren2019,Sauer2021}. Studies have also shown that the RPA captures certain effects of strong electronic correlation \cite{Furche2001,Burke2005,GorlingHesselmann2011a,Gross2012}. Although most calculations are successfully employed non-self-consistently, i.e., the RPA energy is evaluated on top of the ground state of a simpler functional, reaching self-consistency is desirable as it provides an unbiased solution and gives access to interaction-induced changes in the ground-state density. Being diagrammatically equivalent to the $GW$ approximation, the RPA density is expected to be similar to the
$GW$ density \cite{SchefflerCaruso2013,Rinke2015}.  

Self-consistent calculations through the RPA KS potential 
have so far been performed on atoms \cite{Barth2007} and small molecules \cite{GironcoliLinh2014,Gould2019,Gorling2025}. These calculations have demonstrated that the RPA correlation potential exhibits several important features of the exact correlation potential. These include the atomic shell oscillations \cite{Gonze1994,Engel2005,Barth2007}, the $\a/2r^4$ asymptotic tail \cite{Gonze2003}, and the correlation peak that appears when dissociating a covalent bond \cite{Baerends1996,Gross2012,Gould2019}. These features, completely absent in local and semi-local approximations, can be attributed to the dependency of the RPA functional on both the occupied and unoccupied KS orbitals \cite{Kronik2008}. In addition, inter-shell-peaks due to exchange interactions are well described since the exchange term is treated exactly within RPA. The same property is also approximately captured by nonlocal functionals with a dependency on the occupied KS orbitals such as hybrid and meta-GGA (generalized gradient approximation) functionals \cite{Hellgren2014,Perdew2016}.  

Early calculations on semiconductors were only partially self-consistent \cite{Sham1987,Rubio2006}, or employed the quasiparticle approximation \cite{Kresse2014}. Only recently was it shown that these approximations may strongly influence the results, particularly the quasiparticle approximation \cite{Kresse2021}. Nevertheless, these works confirm that the RPA KS potential underestimates the band gap in solids, and that the derivative discontinuity of the exchange-correlation energy is large, corresponding to around 30\%-50\% of the gap \cite{Balduz1982,Rubio2006,MoriSanchez2012}. 
This is not surprising given that the KS potential is designed to reproduce the density of the interacting system \cite{Sham1965}. As such, the KS orbitals and eigenvalues (except for the highest occupied \cite{Barth1985}) do not have a rigorous physical meaning and should, therefore, not be interpreted as excited states of the interacting system. 

On the other hand, the KS system often represents a good starting point for calculating absorption spectra within time-dependent DFT \cite{Gross1996,tddftbook2006} and quasi-particle spectra within many-body perturbation theory \cite{Louie1986,Gunnarsson1998,Reining2018}. Since RPA and $GW$ share the same total energy functional, it is possible to show that the RPA KS band structure provides an optimal starting point for perturbative $G_0W_0$ band structure calculations \cite{Gonze2004}. At the same time, the $G_0W_0$ gap becomes equal to the RPA gap, calculated from the derivative of the RPA total energy with respect to the number of particles. The RPA potential is thus an interesting alternative to fully self-consistent $GW$, which is technically difficult to achieve due to the non-Hermitian and frequency-dependent $GW$ self-energy. Indeed, only few fully self-consistent $GW$ results are available \cite{vanleeuwen2006,Kotliar2012,Kutepov2016,Kutepov2017,Kresse2018}. Most calculations employ various quasi-self-consistent approaches: self-consistency through the Green's function only, keeping the screened interaction fixed \cite{Kresse2007}, quasiparticle self-consistency via a static approximation to the $GW$ self-energy \cite{Faleev2006}, and quasi-self-consistent approaches based on the simultaneous optimization of a parameter-dependent Hubbard \cite{Giustino2012} or hybrid functional \cite{Scheffler2013,Mitra2013,Wirtz2021}. In this context, the RPA starting point stands out as it requires no approximation from a DFT perspective, at least for the evaluation of the band gap.

The RPA, just like hybrid and meta-GGA functionals, is an implicit functional of the density. There is thus no analytical expression for the exchange-correlation potential, which complicates its numerical evaluation. Instead, in each iteration towards self-consistency, one has to solve an integral equation known as the optimized effective potential (OEP) equation \cite{Horton1953,Shadwick1976}. Although numerically challenging, there exists a number of stable implementations both with plane-waves \cite{Gorling1999,GironcoliLinh2014,Hellgren2024} and localized basis sets \cite{Engel2005,Barth2007,Barth2009,Kronik2009,Gorling2021}. 

In this work, we evaluate the RPA KS potential for a set of solids (C, Si, BN, LiF, MgO, TiO$_2$) by generalizing a plane-wave implementation previously applied to molecules \cite{GironcoliLinh2014}. The RPA KS potential is then used to study starting-point dependencies within RPA and $G_0W_0$. In all cases considered, we find that the RPA total energy has a well-defined minimum, and propose a variational approach for optimizing parameter-dependent hybrid functionals. This approach is compared to our previously proposed method based on quasi-self-consistency for the $G_0W_0$ band gap \cite{Wirtz2021}. Focusing then on the PBE0 hybrid functional, we show that, for these weakly correlated solids, the resulting hybrid KS potentials accurately mimic the RPA KS potentials, yielding similar band structures, lattice constants and RPA band gaps.  

The paper is organized as follows. In Sec.~\ref{sec2}, we review the basic equations of self-consistent RPA and derive two different RPA-based approaches for optimizing the exact-exchange fraction in hybrid functionals. In Sec.~\ref{sec3}, we present the computational setup and discuss results for KS potentials, lattice constants and RPA/$G_0W_0$ gaps. Finally, in Sec.~\ref{sec4}, we give our conclusions.

\section{\label{sec2}Theory and implementation}
In this section, we review the equations for self-consistent RPA and their 
implementation based on an eigendecomposition of the KS linear density response function \cite{Gironcoli2009,GironcoliLinh2014}. We also discuss ways to achieve quasi-self-consistency within RPA while simultaneously optimizing parameter-dependent hybrid functionals.
\subsection{\label{sec2a}Self-consistent RPA}
The RPA correlation energy functional can be derived within many-body perturbation theory as a partial resummation 
of ring diagrams equivalent to those of the $GW$ approximation \cite{klein1961,Walecka2003,SchefflerCaruso2013,Rinke2015}. However, while the diagrams in $GW$ are evaluated with a dressed Green's function, in RPA they are evaluated with an independent-particle Green's function, $G_s$. The final expression can either be written in terms of $G_s$ or in terms of the independent-particle linear density response function, $\chi_s=-iG_sG_s$. 
In the latter case, we can write
\be
E^{\rm RPA}_{\rm c}=\int_0^\infty \!\frac{d\w}{2\pi}\Tr\!\!\left\{\ln[1-v\chi_s(i\w)]+v\chi_s(i\w)\right\}
\label{rpac},
\ee
where $v$ is the Coulomb interaction, and $\Tr\{ AB \}=\int d \br d \br' A(\br,\br')B(\br',\br)$. Combined with the exact-exchange (EXX) energy functional, the RPA provides a well-defined 
orbital-dependent exchange-correlation (xc) functional within the KS DFT framework. Although most calculations are done on top of the 
ground state of another functional, i.e., $\chi_s$ is constructed from the eigenvalue spectrum of, e.g., PBE (Perdew-Burke-Ernzerhof) \cite{Ernzerhof1996,Ernzerhof1997} or LDA (local density approximation), full 
self-consistency, via a local xc potential, $v_\xc=\d E_\xc/\d n$, can also be achieved. However, since the dependence on the density is only implicit, the functional derivative of the xc energy with respect to the density must be evaluated using the chain rule through the total KS potential, $V_s=v_{\rm ext}+v_{\rm H} + v_\xc$, where '$\rm ext$' and '$\rm H$' stand for external and Hartree, respectively. This implies solving the OEP equation \cite{Horton1953,Shadwick1976} 
\be
\int d\br'\chi_s(\br,\br',\w=0)v_{\rm xc}(\br')=\frac{\d E_{\rm xc}}{\d V_s(\br)}.
\label{lss}
\ee
The OEP equation is well-known from EXX theory, where expressions for the right hand side can be found in many works (see for example Refs. \cite{Iafrate1992,Perdew2003b}). The RPA correlation contribution can be evaluated as $\d E^{\rm RPA}_{\rm c}/\d \chi_s\times\d \chi_s/\d V_s$ \cite{Gonze2003}, giving 
\be
\frac{\d E_{\rm c}}{\d V_s(\br)}=-\int_0^\infty \!\frac{d\w}{2\pi}\Tr\!\!\left\{\frac{vD(\br,i\w)v\chi_s(i\w)}{1-v\chi_s(i\w)}\right\},
\label{dedv}
\ee
where 
\be
D(\br,i\w)=\frac{\d \chi_s(i\w)}{\d V_s(\br)}.
\ee
An expression for $D$ in terms of KS orbitals and eigenvalues is given in Appendix A. 

Implementations of the OEP equation for RPA exist for atoms in a spline basis set \cite{Barth2007,Barth2010}, for molecules within Gaussian basis sets \cite{Gorling2025}, and for solids within the PAW (projected
augmented wave) framework \cite{Kresse2021}.
In this work, we start from an existing plane-wave implementation within the ACFDT (adiabatic connection fluctuation dissipation theory) package of the \verb|Quantum ESPRESSO| (QE) distribution \cite{Gironcoli2009,GironcoliLinh2014,Giannozzi2017}. In this implementation, $v\chi_s$, or more precisely $v^{1/2}\chi_sv^{1/2}$, is decomposed in terms of its eigenvalues $v_{\b}$
and eigenvectors $V^{\b}$. 
The lowest eigenvalues, at every momentum ${\bf q}$ and imaginary frequency $i\w$, are 
obtained by iterative diagonalization of the generalized eigenvalue problem 
\be
\chi_s({\bf q},i\w)|V_{\bf q}^{\b}\ket=\nu_\b({\bf q},i\w) v_{\bf q}^{-1}|V_{\bf q}^{\b}\ket.
\label{gep}
\ee
The action of $\chi_s$ on a trial eigenvector (or "eigenpotential") is nothing but a density response, which can conveniently be determined within density functional perturbation theory (DFPT) \cite{Testa1987,Giannozzi2001}. 

With the eigendecomposition of $\chi_s$, the correlation energy in Eq.~(\ref{rpac}) reduces to a simpler expression 
\be
E_{c}=\frac{1}{N_{\bf q}}\sum_{\bf q}\int_0^\infty \!\frac{d\w}{2\pi}\sum^{N_{\nu}}_{\rm \b}\ln[1-\nu_\b({\bf q},i\w)]+\nu_\b({\bf q},i\w),
\label{en_eig}
\ee
where $N_{\nu}$ is the number of eigenvalues. It is well-known that the response function and the correlation energy converge slowly with respect to the number of unoccupied states \cite{blugel2012,blugel2013,blugel2015}.  Although the correlation energy in Eq.~(\ref{en_eig}) also converges rather slowly with respect to $N_{\nu}$, we have observed that energy differences are usually well-converged already at 10$\times N_{e}$ \cite{Gironcoli2018,Baguet2021}, where $N_e$ is the number of electrons. The derivative (see (Eq.~(\ref{dedv})) is, similarly, given by
\be
\frac{\d E_c}{\d V_s(\br)}=-\frac{1}{N_{\bf q}}\sum_{\bf q}\int_0^\infty \!\frac{d\w}{2\pi}\sum^{N_{\nu}}_{\rm \b}\frac{D^\b_{\bf q}(\br,i\w)\nu_\b({\bf q},i\w)}{1-\nu_\b({\bf q},i\w)},
\ee
where
\bea
D^\a_{\bf q}(\br,i\w)&=&2\sum^{\rm occ}_{{\bf k}n}\vf^*_{n\bf k}(\br)\d\vf^1_{n\bf k}(\br)+c.c.\nn\\
&&-2\sum^{\rm occ}_{{\bf k}n}\vf^*_{n\bf k+q}(\br)\d\vf^2_{n\bf k+q}(\br)+c.c.\nn\\
&&-2\sum^{\rm occ}_{{\bf k}nn'\s} \bra\D^\a\vf^{-\s}_{n{\bf k+q}}|\D^\a\vf^{\s}_{n'{\bf k+q}}\ket\nn\\
&&\,\,\,\,\,\,\,\,\,\,\,\,\,\,\,\,\,\,\,\,\,\,\times\vf_{n{\bf k}}(\br)\vf_{n'{\bf k}}^*(\br)\nn\\
&&+2\sum^{\rm occ}_{{\bf k}n\s}\D^\a\vf^{\s}_{n{\bf k+q}}(\br)[\D^\a\vf^{-\s}_{n{\bf k+q}}(\br)]^*
\label{der0}
\eea
and
\be
|\D^\a\vf^{\s}_{n{\bf k+q}}\ket=\sum_m^{\rm unocc}|\vf_{m{\bf k+q}}\ket\frac{\bra \vf_{m{\bf k+q}}|V^\a_{\bf q}|\vf_{n{\bf k}}\ket}{\s\w+\ve_{n{\bf k}}-\ve_{m{\bf k+q}}}
\label{der1}
\ee
\be
|\d\vf^1_{n\bf k}\ket=\sum_m^{\rm unocc}|\vf_{m{\bf k}}\ket\frac{\bra \vf_{m{\bf k}}|V^\a_{\bf -q}|\D^\a\vf_{n{\bf k+q}}\ket}{\ve_{n{\bf k}}-\ve_{m{\bf k}}}
\label{der2}
\ee
\bea
|\d\vf^2_{n\bf k+q}\ket&=&\sum_{m}^{\rm unocc}\sum_{m'\s}^{\rm occ}|\vf_{m{\bf k+q}}\ket\frac{\bra \vf_{m{\bf k+q}}|\D^\a\vf^{\s}_{m'{\bf k+q}}\ket}{\ve_{n{\bf k+q}}-\ve_{m{\bf k+q}}}\nn\\
&&\,\,\,\,\,\,\,\,\,\,\,\,\,\,\,\,\,\,\,\,\,\,\times\bra \vf_{m'{\bf k}}|V^\a_{{-\bf q}}|\vf_{n{\bf k+q}}\ket.
\label{der3}
\eea
The quantities in Eqs.~(\ref{der1}-\ref{der3}) can be determined using DFPT. Furthermore, since $\d E_{\rm x}/\d V_s$ is equal to the density response of the Fock exchange potential and the left-hand side of Eq.~(\ref{lss}) is equal to the density response of $v_\xc$, DFPT can also be used to evaluate the terms in Eq. (\ref{lss}). Combined with an iterative approach for solving the OEP equation \cite{Perdew2003a,GironcoliLinh2014}, this computational scheme was used to generate accurate RPA xc potentials for molecules \cite{GironcoliLinh2014}. The same approach was also successfully employed to generate EXX potentials for solids \cite{Wirtz2021}. In this work, we use a modified version of the ACFDT package of QE and generalize the implementation of the RPA potentials to solids.

The RPA total energy functional has a minimum at the OEP stationary point, provided the static density response function within self-consistent RPA has negative eigenvalues \cite{Barth2006}. This cannot be proven in general but no counter example has so far been found. The variational property of RPA  can then be used to optimize computationally cheaper parameter-dependent approximations such as hybrid functionals, through a constrained minimization. For example, in the PBE0$\a$ functional, the fraction (\%) of exact exchange, $\a$, included within PBE0 \cite{Barone1999}, is allowed to vary. By evaluating the RPA total energy on top of OEP-PBE0$\a$ KS orbitals (an approximation from now on denoted OEP$\a$ to distinguish it from the more "standard" PBE0$\a$ functional that uses a nonlocal exchange potential), we can locate the optimal $\a$ value by setting
\be
\frac{d E[V_s^{\rm OEP \a}]}{d\a}=0.
\label{rpaopt}
\ee
The RPA-optimized hybrid KS potential, $V_s^{\rm OEP\a_{RPA}}$, is expected to yield KS orbitals and eigenvalues similar to the self-consistent RPA potential, $V_s^{\rm RPA}$. This approach was tested on molecules in Refs.~\cite{Baguet2021,Casula2022,Hellgren2024}  and shown to give reasonable values for $\a$.
\begin{figure}[t]
\center
\includegraphics[width=0.95\linewidth]{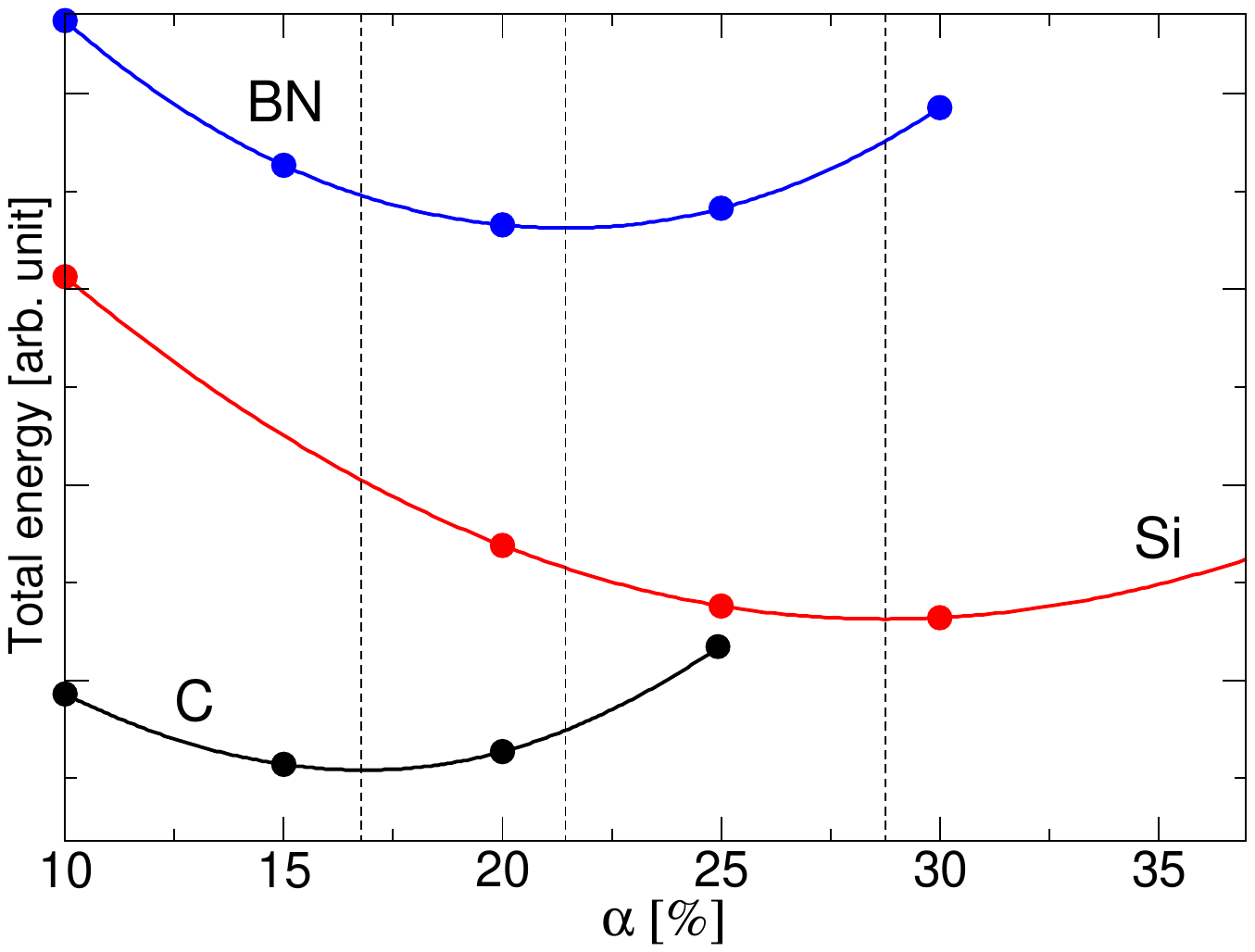}
\caption{RPA total energy of C, BN, and Si, evaluated on top of an OEP$\a$ hybrid functional density, plotted as a function of $\a$. The minima are marked with dashed vertical lines. The curves have been vertically shifted to fit on the same plot.} \label{alfaopt}
\end{figure}
\begin{figure}
\center
\includegraphics[width=1.0\linewidth]{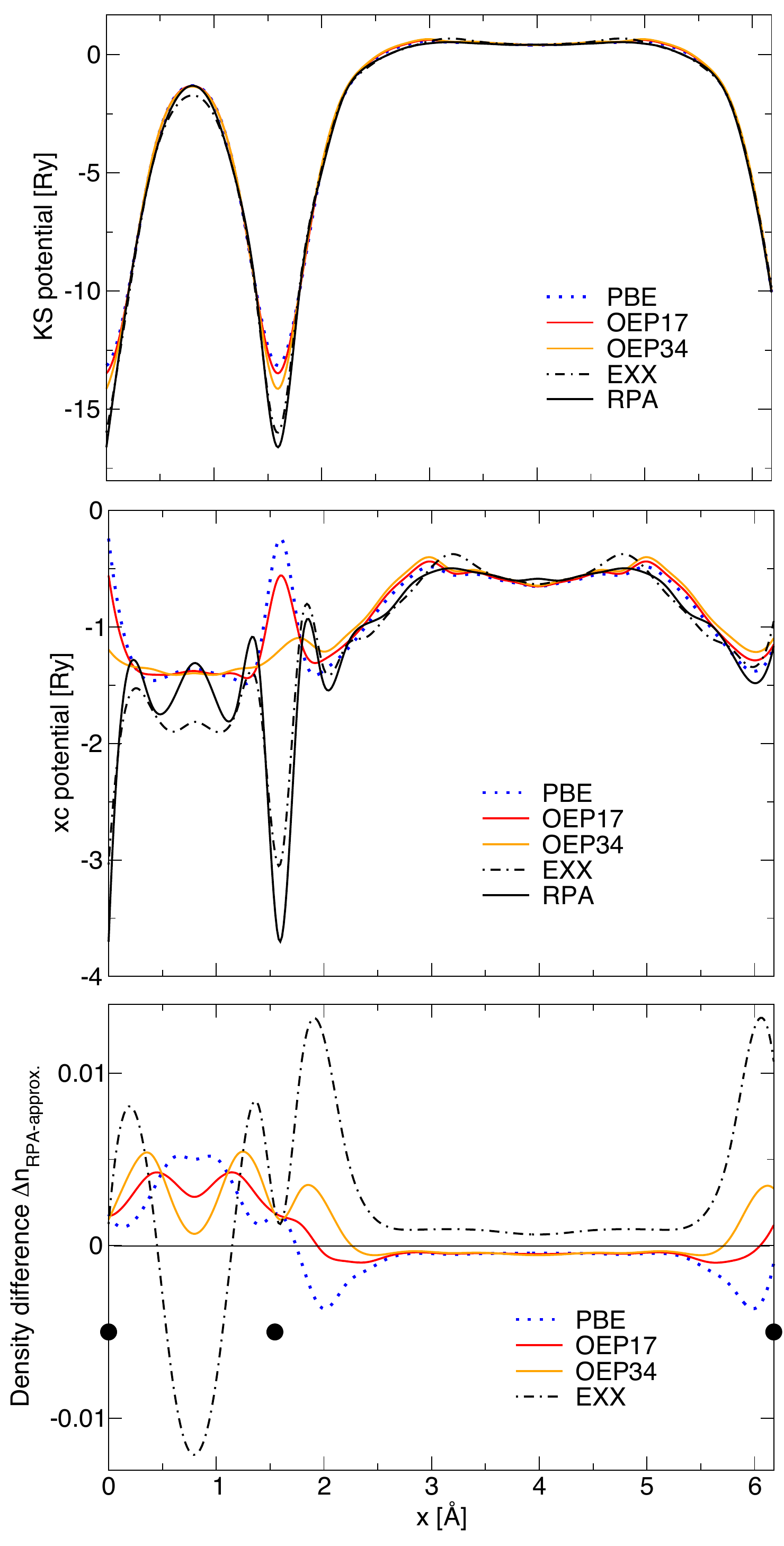}
\caption{KS potentials and densities of C along the [111] direction. Top panel: total KS potential calculated using PBE, EXX, RPA, optimized OEP17, and OEP34. Middle panel: the xc part of the KS potentials. Bottom panel: density difference $\D n_{\rm RPA-approx.}$, where "approx." corresponds to PBE, EXX, OEP17, and OEP34. The positions of the carbon atoms are marked with black dots.} \label{pot_c}
\end{figure}
\subsection{\label{sec2b}RPA band gap}
The RPA is an approximation to the ground-state energy. The RPA band gap, defined as the difference between the ionization energy and the affinity, $E_g=I-A$, can be obtained from 
the derivative of the RPA ensemble energy with respect to the number of particles \cite{Gonze2004,GrossHellgren2012}. 
This derivative is discontinuous at integer particle numbers; the derivative on the left-hand side equals the negative of the ionization energy
\be
-I=\ve_{v}+\bra v|\S_s^{ GW}(\ve_{v})-v^{\rm RPA}_{\rm xc}| v \ket
\label{ion}
\ee 
and the derivative on the right-hand side equals the negative of the affinity 
\be
-A=\ve_{c}+\bra c|\S_s^{ GW}(\ve_{c})-v^{\rm RPA}_{\rm xc}| c \ket.
\label{aff}
\ee 
The $GW$ self-energy appears here, confirming the close connection between RPA and $GW$. The labels $v$ and $c$ denote the RPA KS valence and conduction band states, respectively, and $v^{\rm RPA}_{\rm xc}$ is the xc part of the RPA KS potential. Equations ~(\ref{ion}) and (\ref{aff}) are valid only when the OEP equation (Eq.~(\ref{lss})) is fulfilled. These equations are identical to the standard expressions used in $G_0W_0$ calculations based on a KS reference system \cite{Gonze2004,Kresse2014,Wirtz2021}, except for the renormalization factor that is usually multiplied to the matrix element.  

Analogous formulas can be derived for PBE0$\a$ evaluated with an OEP$\a$ ground state. The $GW$ self-energy is replaced by the nonlocal hybrid potential, which we here denote $\S_s^{\rm PBE0\a}$, and the RPA potential is replaced by the local hybrid xc potential $v^{\rm OEP\a}_\xc$, obtained as the solution to the corresponding hybrid OEP equation.
In Ref.~\cite{Wirtz2021}, this similarity was used to evaluate Eqs.~(\ref{ion}) and (\ref{aff}) approximately by replacing $v^{\rm RPA}_{\rm xc}$ with 
 $v^{\rm OEP\a}_\xc$. Enforcing the RPA gap to equal the gap of the OEP$\a$ hybrid functional yields
\bea
\bra c|\S_s^{ GW}(\ve_{c})-\S_s^{\rm PBE0\a}| c \ket-\,\,\,\,\,\,\,\,\,\,\,\nn&&
\\\bra v|\S_s^{ GW}(\ve_{v})-\S_s^{\rm PBE0\a}| v \ket &=&0,
\label{lopt}
\eea
where the valence and conduction states are now given by OEP$\a$. In addition to generating an approximate starting point for $G_0W_0$ calculations, this scheme provides an alternative approach for optimizing the $\a$-parameter. 
The optimal fraction of exact exchange fulfilling Eq. ~(\ref{lopt}) will from now on be denoted $\a_{\rm opt}$ to distinguish it from the value $\a_{\rm RPA}$ obtained by fulfilling Eq.~(\ref{rpaopt}). Having access to the RPA potential, we will, in the next section, compare the results obtained from these two different approaches to the results obtained using the true RPA potential.  

\section{\label{sec3}Applications}
In this section, we present the computational setup and show results obtained on a set of solids (C, Si, BN, LiF, MgO, rutile-TiO$_2$). We determine the self-consistent RPA potential and the RPA-predicted optimal fraction of exact exchange within the PBE0$\a$ hybrid functional, using the two approaches described in Sec.~\ref{sec2}. The performance of the RPA and the optimized hybrid functionals is evaluated by looking at densities, KS band structures, lattice constants, and RPA/$G_0W_0$ gaps.  

\begin{table*}[t]
\caption{\label{tab1} KS band gaps (eV) of C, Si, BN, LiF, MgO, and rutile TiO$_2$, between $\G$ and different k-points. Results for EXX and RPA are compared to previous calculations reported in the literature. The approximation OEP$\a_{\rm RPA}$ corresponds to the RPA optimized hybrid functional, 
with the optimal value for $\a$ (\%) given in the preceding column. The MAE/MARE are determined with respect to the RPA result obtained in this work. Results marked with (*) or ($^{\dagger}$) are the approximate RPA calculations from Ref. \cite{Rubio2006} and Ref. \cite{Kresse2014}, respectively.}
\begin{ruledtabular}
\begin{tabular}{l l c c c c c c c }
Solid   &k  & PBE &EXX  & EXX lit.\footnote[1]{Ref. \cite{Kresse2014}.}& RPA & RPA lit.\footnote[2]{Results for C, Si, BN and MgO are from Ref. \cite{Kresse2021}. Results for LiF are from Ref. \cite{Rubio2006} (*) and Ref. \cite{Kresse2014} ($^{\dagger}$).}&$\a^{\rm RPA}$&OEP$\a_{\rm RPA}$\\
\hline
C &   & &     &  & &&& \\
&  $\G$& 5.60 &6.20   &6.20  & 5.76&5.75&17 &5.67\\
&  L   & 8.47 &9.07   &9.07  & 8.63&8.62&&    8.54\\
&  X   & 4.81 & 5.36  &5.36  & 4.89&4.89&&    4.84 \\
Si &   &	& &  & &&& \\
&  $\G$& 2.56 &3.18   &  3.15 &2.71 &2.68&29&2.68 \\
&  L   & 1.52 &2.21   &  2.29 &1.62 &1.64&  &1.66 \\
&  X   & 0.69 &1.37   &  1.35 &0.80 &0.74&  & 0.78\\
BN &   &	& &  & &&& \\
&  $\G$&  8.80  &9.72 &  9.80  & 9.01  &  9.12 &  22  & 8.96  \\
&  L   & 10.19 & 11.07 & 11.19 & 10.32 &10.37 &&        10.34 \\
&  X   &  4.54&  5.57 &   5.57 & 4.73  &  4.70 & &       4.67  \\
LiF &  &  &   &  & &&& \\
&  $\G$& 9.25 & 11.28   & 11.33   & 9.81& 9.5*/10.21$^\dagger$&26&  9.75 \\
&  L   & 10.92 &13.08   & 13.11   &11.58 &11.3*/11.86$^\dagger$&    &11.47 \\
&  X   & 14.95 &16.97   & 17.02   &15.65 &15.6*/15.88$^\dagger$&    &15.46\\
MgO &  &  &   &   & &&& \\
&  $\G$& 4.77 & 6.60   & 6.55  &5.21 &5.23&26 &5.21 \\
&  L   & 7.93 & 9.64   & 9.71  &8.30 &8.36&    &8.32 \\
&  X   &9.19  &10.88  & 10.88 & 9.55 &9.50&    &9.56 \\
TiO$_2$ &  &  &   &   & &&& \\
&  $\G$& 1.91 &4.31  &-  &2.40 &- &25  &2.44\\
\hline
MAE&     &0.30  &0.97  &  & &&&0.06 \\
MARE(\%)&     &5.8  &22.1  & & &&&1.1 \\
\end{tabular}
\end{ruledtabular}
\end {table*}
\begin{figure*}
\center
\includegraphics[width=1.0\linewidth]{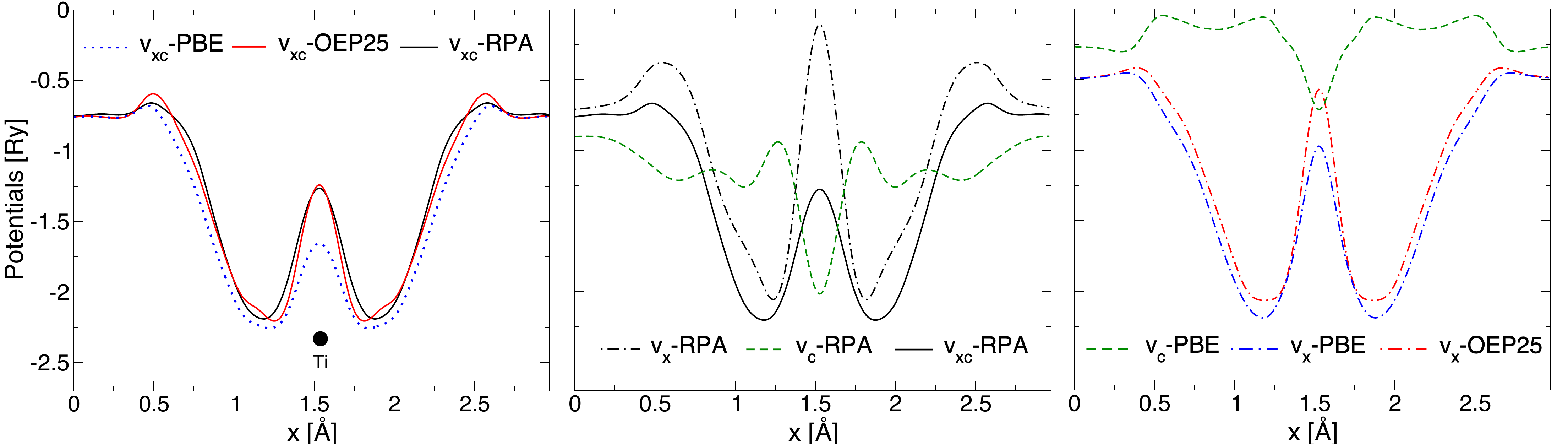}
\caption{KS potentials of TiO$_2$ along the [001] direction. Left panel: xc potentials calculated with PBE, RPA, and the optimized hybrid functional OEP25. The position of the Ti atom is marked with a black dot. Middle panel: RPA xc potential decomposed into its exchange and correlation components. Right panel: PBE correlation potential and PBE and OEP25 exchange potentials.} \label{pot_tio2}
\end{figure*}
\begin{figure}[t]
\center
\includegraphics[width=1.01\linewidth]{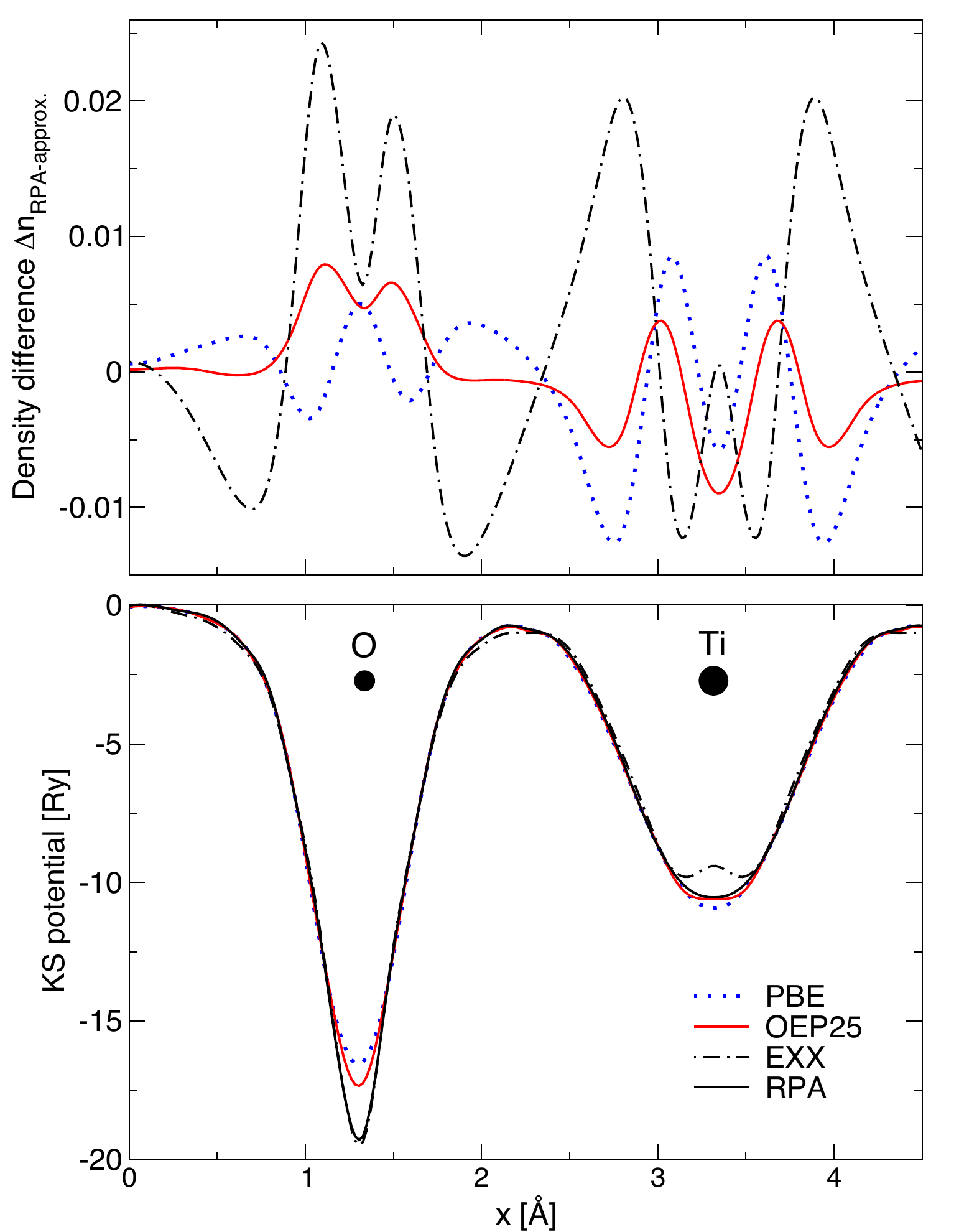}
\caption{KS densities and potential of TiO$_2$ along the [110] direction. Top panel: density difference $\Delta n_{\rm RPA-approx.}$, where "approx." corresponds to PBE, EXX, or OEP25. Bottom panel: total KS potential of TiO$_2$. The positions of the O and Ti atoms are marked with black dots. } \label{totpot_tio2}
\end{figure}
\subsection{Computational details}
Most calculations were performed within the QE software. We used PBE optimized norm-conserving Vanderbilt (ONCV) pseudopotentials \cite{Hamann2013} throughout the work. Lattice constants were determined within 0.002 $\rm \mathring{A}$ using a plane-wave cutoff of 110 Ry and a 8$\times$8$\times$8/6$\times$6$\times$6(shifted) k-point grid for PBE0/RPA, respectively. A convergence plot with respect to the ratio $N_\nu/N_e$ can be found in Appendix B. KS band gaps are converged within $\pm$0.01 eV with a plane-wave cut-off of 80 Ry and a 6$\times$6$\times$6 k-point grid. The RPA KS gaps converged quickly with the number of eigenvalues in the response function, approximately 5$\times N_e$. The frequency integrals in Eqs.~(\ref{dedv}) and (\ref{en_eig})  were converged on a system-adapted grid of 8 frequency points (see Ref. \cite{Gironcoli2018} for details). Convergence plots for KS gaps, lattice constants and $\a_{\rm RPA}$-optimal fractions can be found in Appendix B. The RPA KS gap of the rutile phase of TiO$_2$ was converged within 0.05 eV using 360 eigenvalues and a $3\times3\times3$ k-point grid. 

For the RPA/$G_0W_0$ band gap calculations at experimental lattice constants, we used the YAMBO code with full frequency integration \cite{yambo1,yambo2}. 
Gaps were converged with respect to the number of bands, frequency points and $G$-vector cutoff for the response function. We used a k-point grid of 6$\times$6$\times$6, except in the case of  TiO$_2$ for which we used a 4$\times$4$\times$4 grid. We needed 1000-1500 frequency points and a response function cutoff within 15-26 Ry. The number of bands included were: 160 for C, 110 for Si, 160 for BN, 350 for LiF, 520 for MgO, and 2000 for TiO$_2$. The setup for TiO$_2$ is very close to the one used in a recent precision benchmark calculation \cite{Gonze2024}.  
\subsection{RPA KS potential}
In Table~\ref{tab1}, we present KS band gaps obtained with PBE, EXX, RPA, and the variationally optimized OEP$\a_{\rm RPA}$ hybrid functional. The RPA and EXX results are compared to the results obtained with the PAW method, as implemented in the VASP code \cite{Kresse2021}. For both EXX and RPA, we find good agreement between the two implementations. Differences are less than 0.1 eV, and, in most cases, below 0.05 eV. Variations of similar order are already found at the PBE level (not shown in the Table), suggesting that the differences are mostly due to different treatments of the core electrons. We note that the two RPA reference values for LiF were obtained either by evaluating the screened interaction with EXX orbitals in combination with the plasmon pole approximation (*) \cite{Rubio2006}, or by employing the quasiparticle approximation to the $GW$ self-energy ($\dagger$) \cite{Kresse2014}. The latter was later shown to significantly affect the gaps \cite{Kresse2021}. The effect of the plasmon pole approximation remains unclear. 

We find that self-consistency lowers the RPA energy with respect to the use of PBE, EXX, and OEP$\a$ starting points. By restricting the RPA variational freedom to KS potentials from OEP$\a$, we find, in all cases studied, a minimum at a given $\a=\a_{\rm RPA}$. This is demonstrated in Fig.~\ref{alfaopt}, where we plot the RPA total energy for C, Si and BN as a function of $\a$. The value of $\a_{\rm RPA}$, as presented in Tab.~\ref{tab1}, is, as expected, system-dependent. 

By looking at the mean absolute error (MAE) and the mean absolute relative error (MARE(\%)) of the KS gaps in PBE, EXX and OEP$\a_{\rm RPA}$ with respect to RPA, we see that PBE is closer to RPA (MAE: 0.30 eV and MARE\%: 5.8\%) than EXX (MAE: 0.97 eV and MARE\%: 22.1\%). We also see that RPA and OEP$\a_{\rm RPA}$ agree rather well (MAE: 0.06 eV and MARE\%: 1.1\%), suggesting that the hybrid potential is able to mimic the RPA potential, at least in the set of solids studied here. This validates the $\a$-optimization procedure, which we emphasize only indirectly optimizes the potentials via the total energy. The largest difference is found for C and LiF, where the gaps differ by around 0.1 eV. Focusing on C, we found that it is possible to get a better agreement with RPA by increasing the fraction of exact exchange from 17\% to 34\%. The latter value gives the gaps $\G-\G = 5.74$ eV, $\G-{\rm L} = 8.61$ eV and $\G-{\rm X} = 4.86$ eV, which almost coincide with the RPA gaps. In order to investigate why the $\a$-optimization does not find this seemingly better value of $\a$, we have analyzed the corresponding KS potentials and densities. The top and middle panels of Fig.~\ref{pot_c} show the total and the xc part of the C KS potential along the [111] direction. The density difference with respect to the RPA density is shown in the bottom panel of the same Figure. Visually, it appears difficult to reproduce the full spatial structure of the RPA potential with OEP$\a$. The OEP34 looks somewhat better than OEP17 in the C-C bond region but worsens towards the interstitial region. Looking at density differences is somewhat more informative. The optimized OEP17 functional overall better reproduces the RPA density as compared to OEP34, especially the amount of charge density localized in the vicinity of the C-C bond. The OEP34 instead over-localizes the density, an effect that only gets stronger the larger the fraction of exact exchange. The OEP$\a$ potential thus appears not flexible enough to fully reproduce both gaps and densities using the same value of $\a$.

Let us now focus on TiO$_2$, a system relatively sensitive to a fraction of exact exchange in the PBE functional. Looking at the $\G-\G$ gap, we find very good agreement between RPA and OEP$\a_{\rm RPA}$. The xc potentials are plotted in Fig.~\ref{pot_tio2} along the [001] direction, i.e., along an axis that crosses Ti-atoms only. In the left panel, the RPA xc potential is compared to the OEP25 and PBE xc potentials. PBE clearly predicts a wider potential well and underestimates the height of the barrier at the center of the atom. By including 25\% of exact exchange as within the optimized OEP25, it is possible to reproduce most of the features of the RPA potential. The well width and the height of the barrier now almost coincide. A difference is seen at the edges of the well where OEP25 gives a sharper step-structure, as opposed to the narrowing of the well observed in RPA. 

In order to gain further insights, we have decomposed the xc potentials into their exchange and correlation components, shown in the middle and right panels of Fig.~\ref{pot_tio2}.
The correlation potentials in RPA and PBE are seen to be rather different. Except for a dip at the position of the atom, the PBE correlation potential is almost inverted as compared to the RPA correlation potential. This effect appears similar to what has already been observed in atoms \cite{Engel2005}. The PBE correlation potential thus simulates features of the exchange potential, leading to a compensation of errors between exchange and correlation components. 
 
In Fig.~\ref{totpot_tio2}, we have plotted the total KS potential of TiO$_2$ along a direction that crosses both O and Ti atoms. In the top panel, the density difference with respect to RPA is plotted along the same path for PBE, EXX, and OEP25. As expected, the density difference is minimized with the optimized OEP25 functional, especially in the vicinity of the Ti atom and in-between the O and Ti atoms.
\subsection{Lattice constants}
To further evaluate the performance of the RPA optimized hybrid functionals, we have calculated lattice constants for the set of solids studied above. The results, presented in Tab.~\ref{tab2}, are compared to the RPA results obtained in the present work and to those within the PAW framework \cite{Kresse2010,Kresse2020}. PBE and experimental values are also indicated. The latter, extracted from Ref. \cite{Kresse2010}, are corrected for zero-point anharmonic expansion effects in all cases except TiO$_2$. Since self-consistency was found to play an insignificant role in diamond, the RPA lattice constants are obtained by evaluating the RPA energy on top of PBE orbitals. For the self-consistent hybrid functional calculations, we used the nonlocal hybrid potential. The results are therefore denoted PBE0$\a_{\rm RPA}$. Using the local OEP$\a_{\rm RPA}$ hybrid potential did not change the results. 
\begin{table}[t]
\caption{\label{tab2} Lattice constants ($\rm \mathring{A}$) obtained with PBE, RPA, and PBE0$\a_{\rm RPA}$ (see Table \ref{tab1} for $\a^{\rm RPA}$). Results are compared to the reference RPA and experimental results corrected for zero-point anharmonic expansion effects published in Refs.~\cite{Kresse2010} and \cite{Kresse2020}. The experimental extrapolated $T=0$ K results for the lattice constants $a$ and $c$ of rutile TiO$_2$ are taken from Ref.~\cite{Post2007}. The relative (\%) error for each approximation with respect to experiment is illustrated in Fig.~\ref{latt}.}
\begin{ruledtabular}
\begin{tabular}{l c c c c c c c}
Solid & PBE & PBE0$\a_{\rm RPA}$ &RPA&RPA lit.&  Exp.                   \\
\hline
C &3.567&  3.552 &3.567  &3.565 & 3.553\\
Si &5.479 & 5.435  &5.414&5.437 & 5.421\\
BN &3.621&  3.599 &3.617 &-& 3.592\\
LiF &4.062& 4.007 &3.996&3.998 & 3.972\\
MgO &4.255& 4.208  &4.200& 4.210 &4.189  \\
TiO$_2$ $(a)$ &4.640& 4.586  &4.596&-  & 4.586 \\
TiO$_2$ $(c)$&2.971& 2.945  &2.956&  -& 2.952 \\
\end{tabular}
\end{ruledtabular}
\end {table}

\begin{figure}[b]
\center
\includegraphics[width=0.95\linewidth]{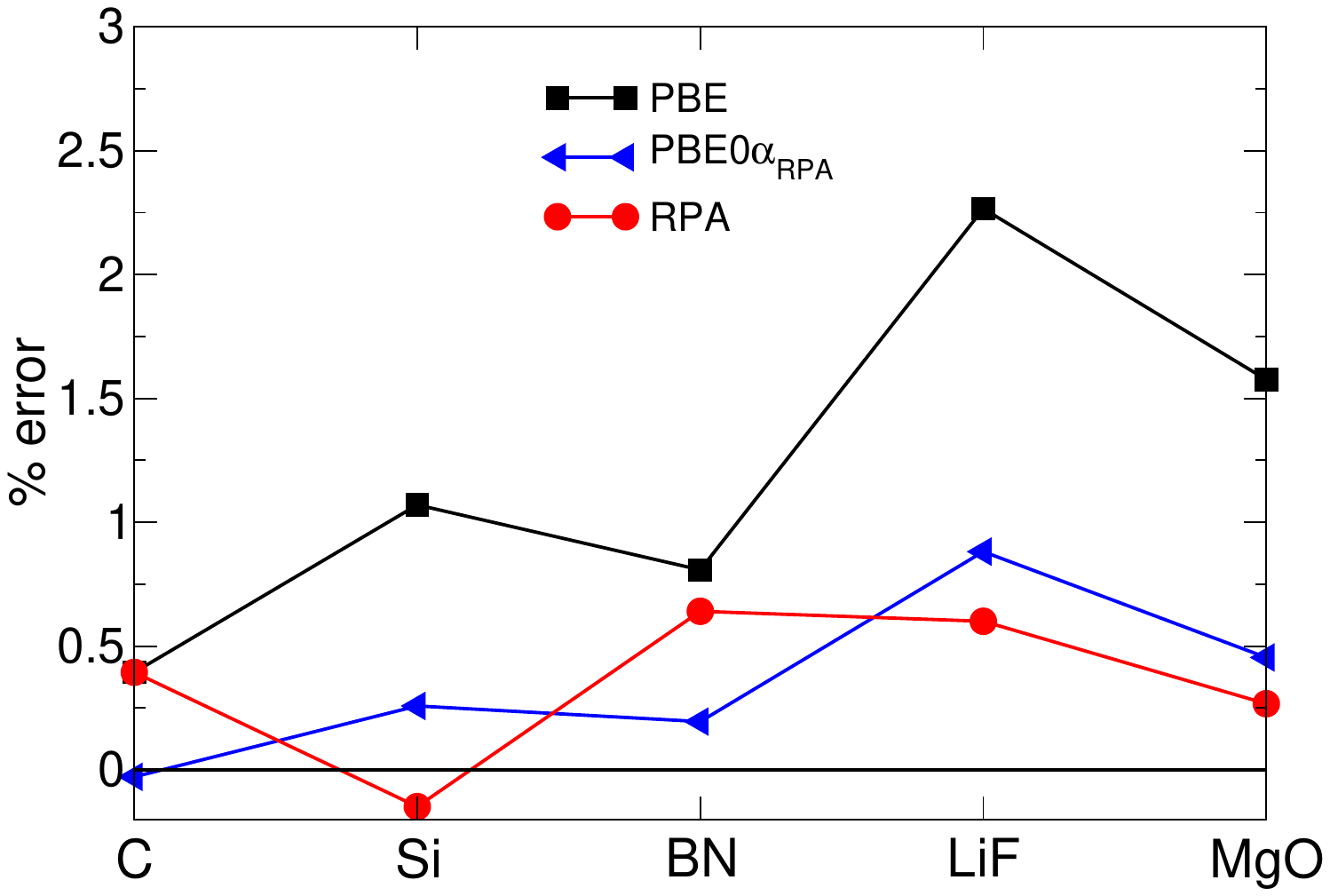}
\caption{The relative (\%) error with respect to corrected experimental results for PBE, RPA, and the optimized hybrid functional PBE0$\a_{\rm RPA}$ (see Table \ref{tab1} for the values of $\a^{\rm RPA}$).} \label{latt}
\end{figure}

The lattice constant is a property obtained from the first derivative of the energy, which is directly determined by the electronic density. 
By comparing PBE0$\a_{\rm RPA}$ and RPA lattice constants, we therefore get a measure of how similar the densities are. The results are illustrated in Fig.~\ref{latt}, where the relative (\%) error with respect to corrected experimental results is plotted for all approximations and solids. For Si, LiF and MgO, a fraction of exact exchange largely improves the PBE lattice constants. At the same time, the PBE0$\a_{\rm RPA}$ results agree well with RPA. Good $\alpha$ values are also found for C and BN, producing lattice constants in better agreement with experimental results than the underlying RPA functional, which, in these cases, gives results very close to PBE. The performance of the optimized hybrid functionals is thus either similar or better than RPA. 

Overall, RPA significantly improves upon PBE, consistent with a conclusion drawn based on results for a larger set of solids \cite{Kresse2010,Kresse2020}. We find good agreement between the RPA lattice constants obtained in this work and those obtained using the PAW method for C, LiF and MgO \cite{Kresse2010,Kresse2020}. In contrast, for Si we see a difference of almost 0.02 $\rm \mathring{A}$, with an underestimation in the present work. However, if we compare to a previous similar calculation using a norm-conserving pseudopotential, we find our result to be overestimated by 0.03 $\rm \mathring{A}$ \cite{Gironcoli2009}. There is thus a 5.38$\rm \mathring{A}$-5.43$\rm \mathring{A}$ span in the published results of the RPA lattice constant of Si. Our result using the ONCV pseudopotential is somewhere in-between previous calculations. This suggests that the results for Si are very sensitive to the construction of the pseudopotential. Further investigations are needed to clear up this issue.

\subsection{\label{sec3c} RPA band gaps}
In Table~\ref{tab2}, we present quasi-particle band gaps evaluated within the $G_0W_0$ approximation. Since we have omitted the renormalization factor in all calculations, we will from now on refer to them as RPA gaps. The difference is relevant since the renormalization factor reduces the values of the gaps by around 0.5 eV. The starting-point dependence is studied by calculating the RPA gaps starting from PBE, OEP$\a_{\rm RPA}$, and RPA KS potentials. In the following, we will indicate the starting-point approximation with the standard notation "$@$". 

A new optimal fraction of exact exchange within OEP$
\a$ was then obtained by the approach detailed in Sec.~\ref{sec2b}. We find that this gap-optimized, $\a_{\rm opt}$, in general, is different from the variationally optimized, $\a_{\rm RPA}$, obtained above. The PBE0$\a_{\rm RPA}$ functional tends to underestimate the gaps except for Si and TiO$_2$, where the gap is overestimated. It is well-known that the PBE0$\a$ gaps are strongly dependent on $\a$. Indeed, if we look at diamond, we see that changing $\a$ from 17\% ($\a_{\rm RPA}$) to 25\% ($\a_{\rm opt}$) increases the PBE0$\a$ gap by about 0.7 eV. In LiF, a change from 26\% ($\a_{\rm RPA}$) to 46\% ($\a_{\rm opt}$) increases the gap by about 2.4 eV. Given these differences between $\a_{\rm RPA}$ and $\a_{\rm opt}$, we conclude that it is difficult to achieve high accuracy on the gap and the density simultaneously.  

\begin{figure}
\center
\includegraphics[width=1.0\linewidth]{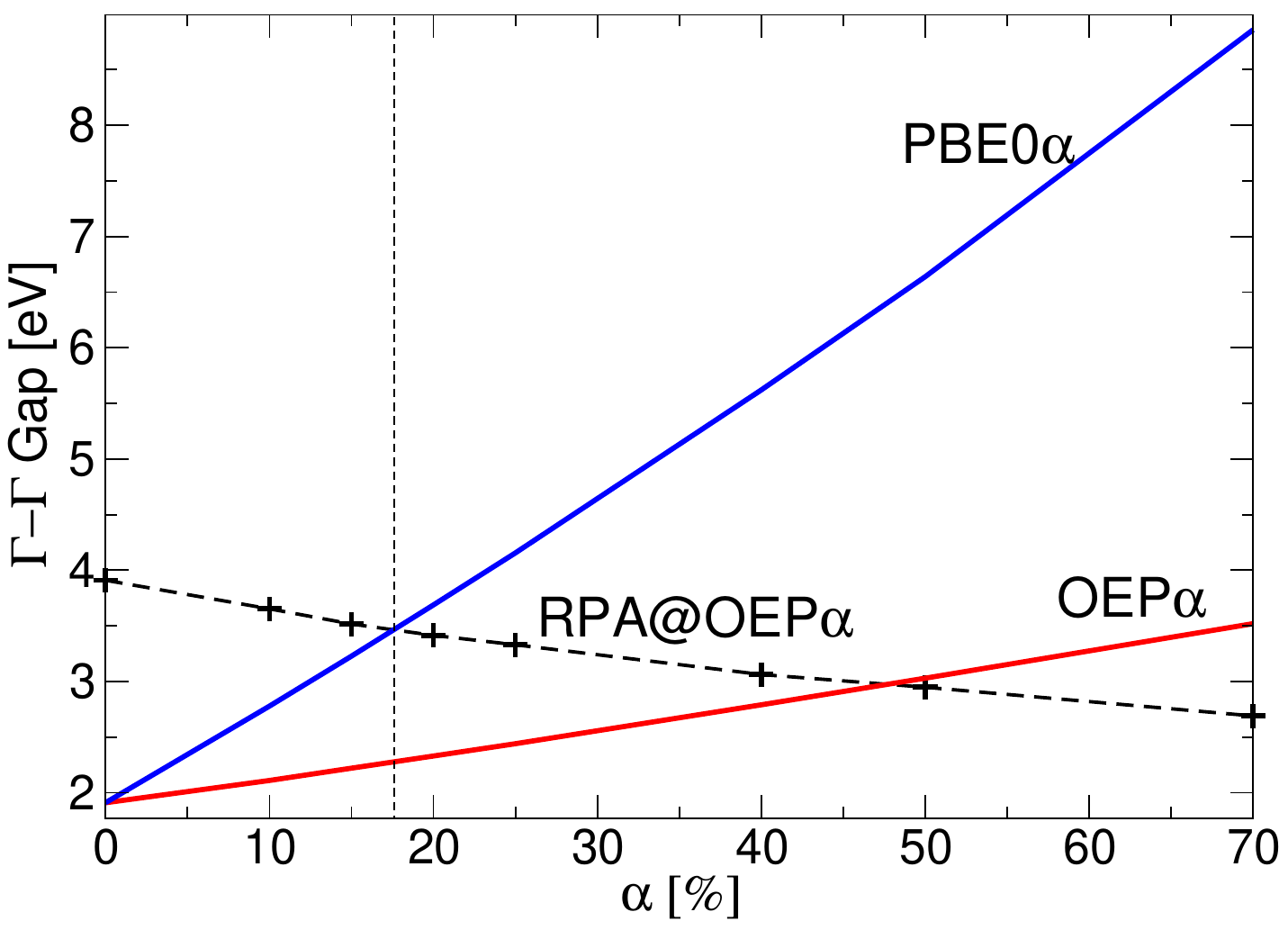}
\caption{$\G-\G$ gap of TiO$_2$ for PBE0$\a$, OEP$\a$, and RPA@OEP$\a$, plotted as a function of $\a$. The crossing of PBE0$\a$ and RPA@OEP$\a$ corresponds to $\a_{\rm opt}$ (marked with a dashed vertical line).} \label{gaptio2}
\end{figure}

For C, Si, and BN, the RPA gaps have a very weak starting-point dependence. In the polar solids LiF and MgO, we see more significant changes, of the order of 0.3 eV between the PBE and RPA starting-points. TiO$_2$ has the strongest starting-point dependence. The gap changes as much as 0.5 eV going from PBE to RPA KS potentials.  
Looking at the relative difference, the change is as large as 15\% in TiO$_2$ but only 2-3\% in MgO and LiF.  The starting-point dependence is also unusual in TiO$_2$. While in most solids the RPA gap increases with a larger fraction of exact-exchange, we see the opposite trend in TiO$_2$ (see Fig.~\ref{gaptio2}). The larger the gap in OEP$\a$ the smaller the RPA gap. With an OEP70 starting point, the gap is as small as 2.7 eV. Similar observations were made for TiSe$_2$, TiS$_2$ and ScN in Ref.~\cite{Wirtz2021} and for TiO$_2$ using an LDA+$U$ starting point in Ref.~\cite{Giustino2012}. With such a strong dependency on the starting point, the only meaningful result is the one obtained with the RPA potential. 
Here, we see that OEP$\a_{\rm opt}$ can provide a similar value.

Next, we compare the RPA gaps to the experimentally measured gaps. In most cases the RPA overestimates the experimental gap. On the other hand, zero-point renormalization due to electron-phonon interactions have recently been shown to reduce the calculated gaps substantially. In Ref. \cite{Kresse2022} the corrections we determined to be -0.323 eV for C, -0.058 eV for Si, -0.400 eV for BN, -1.231 eV for LiF, -0.533 eV for MgO, and -0.349 eV for TiO$_2$. Adding these values to the RPA gaps brings them in better agreement with experiment. MgO and LiF remain, however, underestimated, suggesting the need for vertex corrections.

\begin{table*}[t]
\caption{\label{tab2} Band gaps (eV) of C, Si, BN, LiF, MgO, and rutile TiO$_2$, between $\G$ and different k-points. The RPA gaps (equivalent to $G_0W_0$ gaps without the renormalization factor) are obtained with a KS starting point from PBE, RPA, OEP$\a_{\rm RPA}$ and OEP$\a_{\rm opt}$. The corresponding PBE0$\a$ results and the $\a_{\rm opt}$ (\%) are also given. The $\a_{\rm RPA}$ values are given in Table \ref{tab1}.
}
\begin{ruledtabular}
\begin{tabular}{l l c c c c c c c c}
Solid   &k  & RPA@PBE & RPA@RPA  & PBE0$\a_{\rm RPA}$ & RPA@OEP$\a_{\rm RPA}$ & $\a_{\rm opt}$&PBE0$\a_{\rm opt}$ & RPA@OEP$\a_{\rm opt}$  &Exp.                  \\
\hline 
C &   & &     &  && &&&\\
&  $\G$& 7.67 & 7.73 & 7.04 & 7.69&25&7.72&7.71&7.1\footnote[1]{Ref. \cite{Fuchs1992}.}\\
&  L &  10.63&10.69 & 10.06 & 10.65&& 10.80 &10.67&\\
&  X &  6.25&  6.30& 6.11 &6.29 && 6.72&6.32&\\
Si &   &	& &  & &&&&\\
&  $\G$& 3.39 &3.39  & 4.18 &3.43 &16&3.45&3.43&3.05\footnote[2]{\label{foot2}Ref. \cite{Himpsel1993}.}\\
&  L   & 2.36 & 2.35 & 3.08 &2.39 &&2.37&2.39&\\
&  X   & 1.46 & 1.40 & 2.11 & 1.48   &&1.47&1.49&1.25\footnotemark[2]\\
BN &   &	& &  & &&&&\\
&  $\G$& 11.50 & 11.50 & 10.98 &11.54 &25&11.28&11.53&\\
&  L   & 12.39 & 12.41 &  12.35& 12.44&&12.54&12.43&\\
&  X   & 6.56  &6.57  & 6.36 & 6.62&&6.61&6.61&6.36\footnote[3]{Ref. \cite{Poolton2008}.}\\
LiF &  &  &   &  & &&&&\\
&  $\G$& 14.55 & 14.85 & 12.41 &14.74&46 &14.86 & 14.85&14.20\footnote[4]{Ref. \cite{Olson1976}.}\\
&  L   &  16.25   & 16.54 & 14.12 & 16.42   & & 16.58&16.53 & \\
&  X   & 21.23    &   21.52&  18.49 &  21.44  & & 21.20&21.56&\\
MgO &  &  &   &   & &&&&\\
&  $\G$& 7.72 & 7.98 & 7.29 & 7.85&32&7.90&7.87&7.83\footnote[5]{Ref. \cite{Walker1973}.}\\
&  L   & 11.18 & 11.37 & 10.37 & 11.28&&11.20&11.30&\\
&  X   & 12.17 & 12.98 & 11.53 & 12.27&&12.33&12.28&\\
TiO$_2$ &  &  &   &   & &&&&\\
&  $\G$& 3.91 & 3.40 & 4.16 & 3.33&18&3.50&3.46&3.0\footnote[6]{Ref. \cite{Scheel1996}.}\\
\end{tabular}
\end{ruledtabular}
\end {table*}

\section{\label{sec4}Conclusions}
In this work, we have calculated RPA KS potentials for a set of solids by extending a plane-wave implementation previously applied to molecules. The key feature of this implementation is the eigendecompositon of the KS response function together with the use of DFPT, which eliminates the need to generate unoccupied KS states, both for the energy and the gradient. The computational cost of evaluating the gradient+energy is roughly twice the cost of evaluating the energy alone. We found a smooth convergence of the RPA potential in all cases studied. KS band gaps were converged within $\pm$0.01 eV using a small number of eigenvalues (approximately 5 times the number of electrons). We also found good agreement with first calculations that used the PAW method. 

The RPA total energy was then studied with respect to different starting points. In all solids, self-consistent RPA was lower in energy as compared to RPA with a PBE or OEP$\a$ starting point. By restricting the domain of KS potentials to OEP$\a$ hybrid potentials, a variational method for optimizing $\a$ was developed. We showed that the resulting hybrid potential produced a KS band structure very close to the RPA band structure, and that the corresponding lattice constant was of similar quality as the RPA lattice constant. The method thus provides a reliable tool for generating predictive system-optimized hybrid functionals. Such hybrid functionals can be used to compute properties such as geometries and phonon frequencies, for which RPA may be computationally too demanding. 

The starting-point dependence of the RPA band gaps (equal to the $G_0W_0$ band gaps without the renormalization factor) were then studied together with an alternative approach for optimizing the fraction of exact exchange of the PBE0 functional. For most solids studied here we found a relatively weak dependency on the starting point. An exception is TiO$_2$, where we found variations as large as 1.5 eV  (about 40\% of the gap), depending on the fraction of exact exchange in the OEP$\a$ starting point. Surprisingly, the larger the $\a$ the smaller the RPA gap, similar to what has been found in TiSe$_2$, TiS$_2$, and ScN. 
In general, the optimization of $\a$ via the RPA gap results in values for $\a$ different from those obtained after variational optimization. This demonstrates that, in many cases, it is not possible to get accurate gaps and structures using the same fraction of exact exchange. Which approach is more suitable depends on the property to be studied. It should be noted, however, that the variational optimization is supported by a unique global minimum and, therefore, not limited to the PBE0 functional, but may be applied to more flexible functionals with several parameters to be optimized. Finally, we expect further improvements of the values for $\a$ using the more accurate RPA + exchange (RPAx) functional \cite{Barth2010,Baguet2021}. Such investigations are left for future work. 
\acknowledgements
The work was performed using HPC resources from GENCI-TGCC/CINES/IDRIS (Grant No. A0150914650). 
\section*{Appendix A: $\d\chi_s/\d V_s$}
The functional derivative of $\chi_s$ with respect to $V_s$ is here given in terms of KS orbitals and eigenvalues
\bea
\frac{\d \chi_s(i\w)}{\d V(\br)}&=&\nn\\
&=&\sum_{kk'p}\left\{-\frac{n_{k'}(1-n_{k})(1-n_{p})}{(\ve_p-\ve_{k'})(\pm i\w-(\ve_{k}-\ve_{k'}))}\right.\nn\\
&-&\frac{n_p(1-n_{k'})(1-n_{k})}{(\ve_{k'}-\ve_p)(\pm i\w-(\ve_{k}-\ve_{p}))}\nn\\
&+&\frac{n_k(1-n_{p})n_{k'}}{(\ve_{p}-\ve_{k'})(\pm i\w-(\ve_{p}-\ve_{k}))}\nn\\
&+&\frac{n_k(1-n_{k'})n_{p}}{(\ve_{k'}-\ve_p)(\pm i\w-(\ve_{k'}-\ve_{k}))}\nn\\
&-&\frac{n_{k'}(1-n_{k})n_{p}}{(\pm i\w-(\ve_{k}-\ve_{k'}))(\pm i\w-(\ve_{k}-\ve_{p}))}\nn\\
&+&\left.\frac{n_k(1-n_{k'})(1-n_{p})}{(\pm i\w-(\ve_{k'}-\ve_{k}))(\pm i\w-(\ve_{p}-\ve_{k}))}\right\}\nn\\
&\times&\vf_p(\br)\vf^*_{k'}(\br)\vf_{k}(\br_1)\vf^*_{p}(\br_1)\vf^*_k(\br_2)\vf_{k'}(\br_2)\nn\\
\eea
This expression is equivalent to Eqs.~(\ref{der0})-(\ref{der3}) presented in Sec.~\ref{sec2}.
\begin{figure}[t] 
\center
\includegraphics[width=1.01\linewidth]{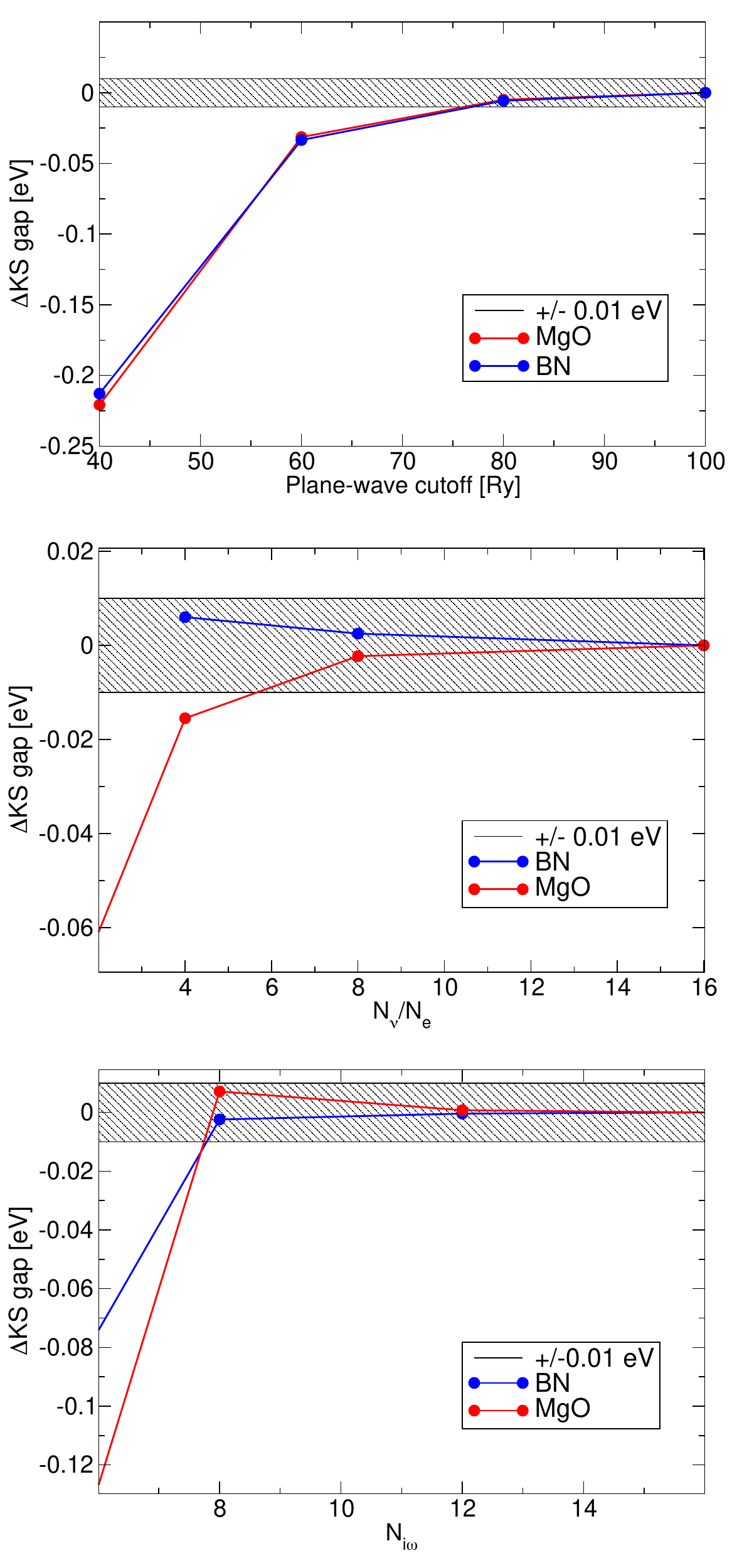}
\caption{Convergence of the KS gap of BN ($N_{\rm e}=8$) and MgO ($N_{\rm e}=16$) with respect to the plane-wave cutoff (top), the ratio $N_{\nu}/N_{ e}$ (middle), and $N_{i\w}$ (bottom). Results are plotted with respect to the most converged calculation. The shaded area corresponds to $\pm 0.01$ eV.} \label{conv_1}
\end{figure}
\begin{figure}[t]
\center
\includegraphics[width=1.01\linewidth]{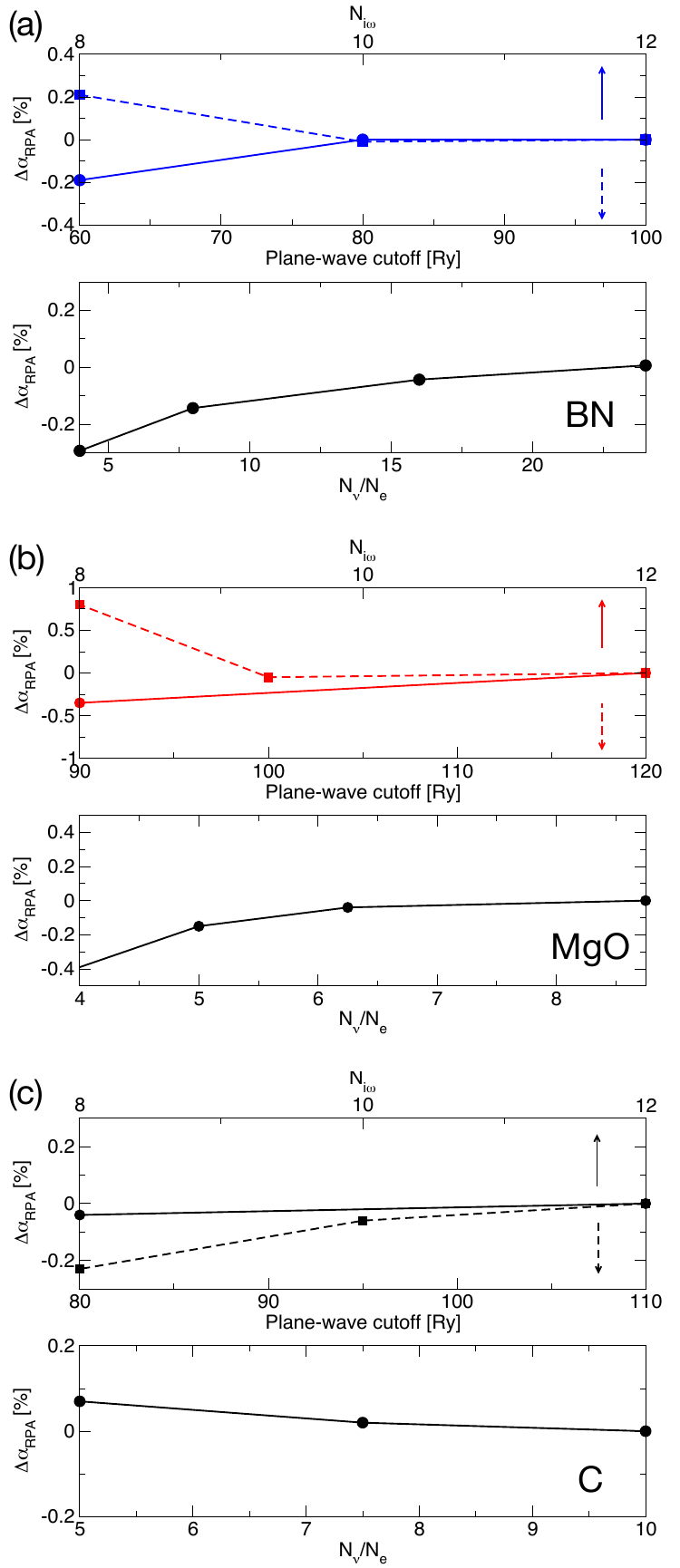}
\caption{Convergence of the optimized fraction of exact exchange $\a^{\rm RPA}$ for (a) BN, (b) MgO, and (c) C with respect to $N_{i\w}$ and the plane-wave cutoff (upper panel), and the ratio $N_{\nu}/N_{e}$ (lower panel). The value of $\a^{\rm RPA}$ is in all calculations extracted from the minimum of the total energy (see Fig. \ref{alfaopt}). Results are plotted with respect to the most converged calculation.
} \label{conv_2}
\end{figure}
\begin{figure}[t]
\includegraphics[width=1.01\linewidth]{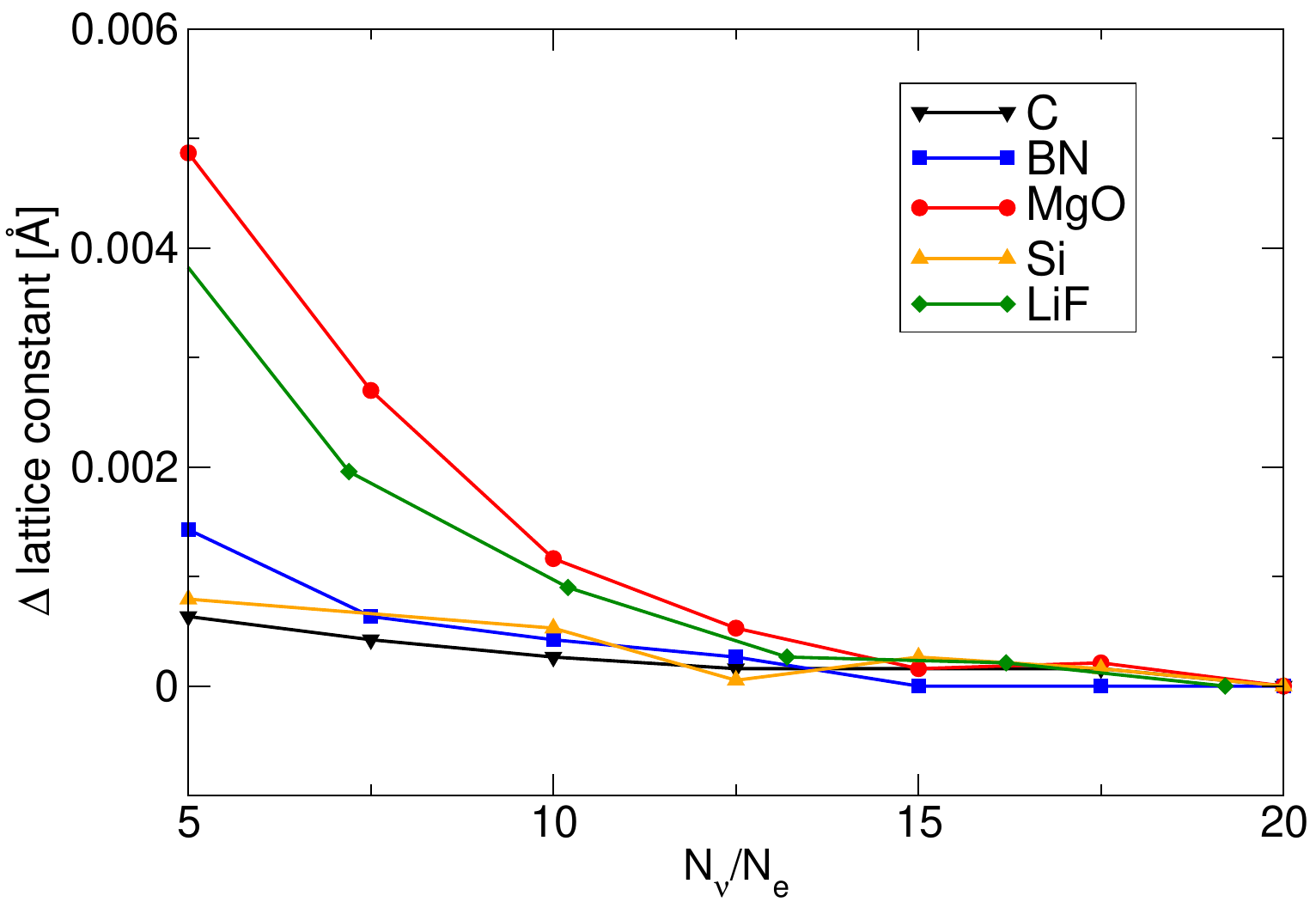}
\caption{Convergence of the lattice constant with respect to the ratio $N_\nu/N_e$ for the different solids studied in this work. Results are plotted
with respect to the most converged calculation.} \label{conv_4}
\end{figure}
\begin{figure}[t]
\center
\includegraphics[width=1.01\linewidth]{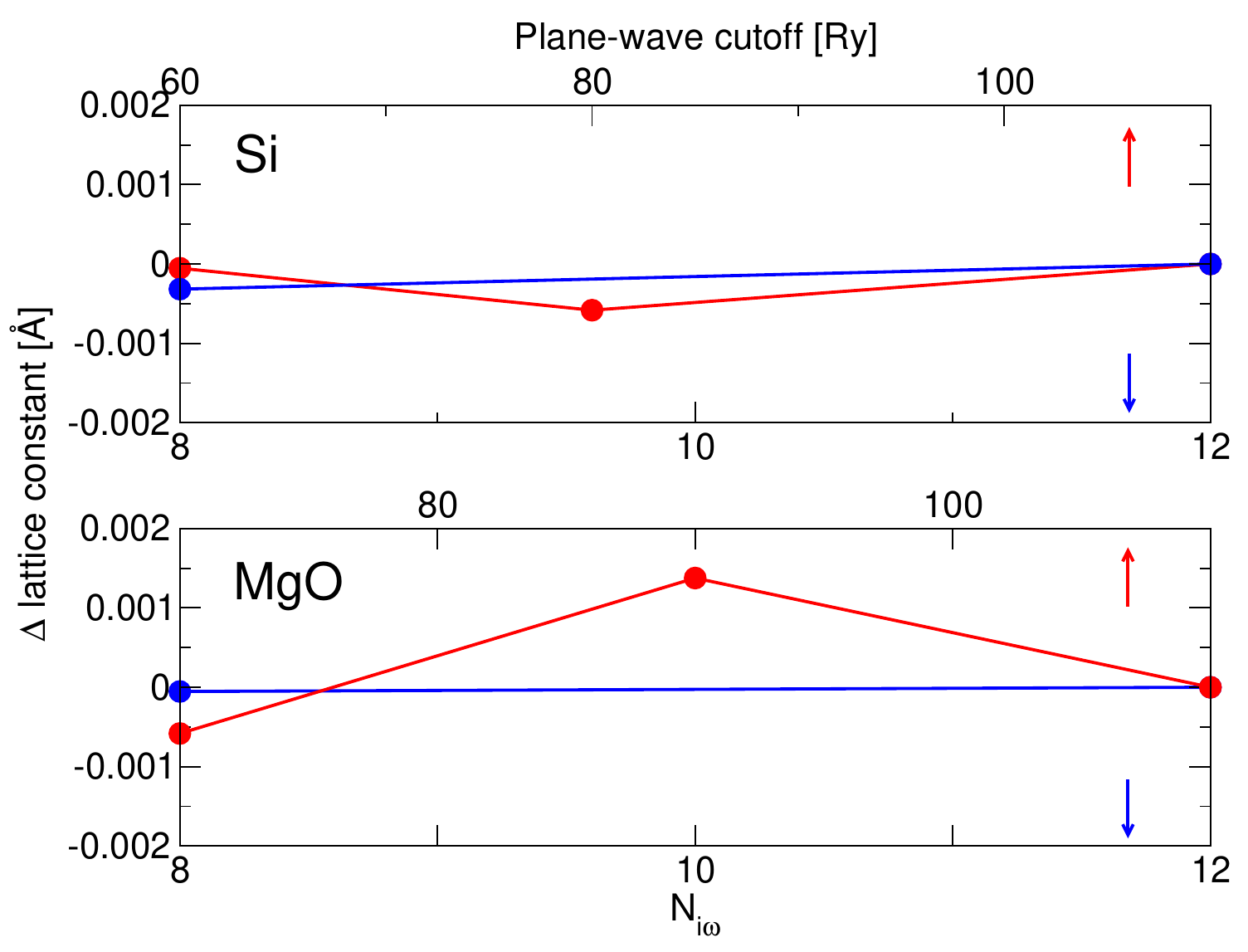}
\caption{Convergence of the lattice constant with respect to the plane-wave cutoff and $N_{i\w}$ for Si and MgO. Results are plotted
with respect to the most converged calculation. } \label{conv_5}
\end{figure}
\section*{Appendix B: Convergence studies}
In this appendix we present convergence studies for some representative cases. 
While the total correlation energy converges rather slowly with respect to $N_{\nu}$ and plane-wave cutoff, the quantities of interest such as KS eigenvalues and the derivative of the total energy converge much faster. Furthermore, with a well-chosen frequency grid \cite{Gironcoli2018}, the integral over $i\w$ converges quickly with the number of grid points ($N_{i\w}$). In Fig. \ref{conv_1}, we plot the convergence of the self-consistent RPA KS gap as a function of the plane-wave cutoff, the ratio $N_{\nu}/N_{e}$, and $N_{i\w}$ for BN and MgO. Our target accuracy, $\pm 0.01$ eV, is marked with a shaded area. With a 80 Ry cutoff, we have reached our target of an error below 0.01 eV. The convergence with respect to the number of eigenvalues, $N_{\nu}$, is smooth and well below an error of 0.01 eV at 8$\times N_e$, which implies 64 and 128 eigenvalues for BN and MgO, respectively. The frequency integral is well-converged at 8 frequency points. 

For extracting the optimized $\a_{\rm RPA}$, it is necessary to converge the minimum of the total energy as a function of $\a$, where the latter refers to the fraction of exact exchange used for generating the OEP$\a$ starting orbitals. The location of the minimum of the curve is found to be highly stable as a function of the ratio $N_{\nu}/N_{e}$. We also find quick convergence with respect to plane-wave cutoff and $N_{i\w}$. In Fig. \ref{conv_2}
we look at the convergence of these parameters for BN, C, and MgO. With an interest of a value for $\a$ within $\pm 1$\%, results are well-converged even with a rather crude setup.

Next, we look at lattice constants. In Fig.~\ref{conv_4} we plot the convergence of the lattice constant as a function of the ratio  $N_{\nu}/N_{e}$ for C, BN, MgO, Si and LiF. A very quick convergence is found for C and Si. Already at 5$\times N_e$, the lattice parameter is converged below 0.001 $\rm \mathring{A}$. The same accuracy is achieved with 10$\times N_e$ for the other solids. Nevertheless, an accuracy of less than 0.06\% is always obtained already at 5$\times N_e$, which is sufficient for most applications. In Fig.~\ref{conv_5} we plot the convergence of the lattice constant as a function of $N_{i\w}$ and the plane-wave cutoff for Si and MgO. As for the other properties, 8 frequency points are sufficient and a rather quick convergence is found with respect to plane-wave cutoff. 
\newpage
%
\end{document}